\documentclass[prl,twocolumn,amsmath,amssymb,superscriptaddress]{revtex4}

\usepackage{amssymb}
\usepackage{graphicx,amsmath}
\usepackage{bm}
\usepackage{times}

\newcommand{\NLij}{i\kern -0.08em j}

\newcommand{\ket}[1]{\lvert{#1}\rangle}

\begin{document}

\title{Dressed-state amplification by a superconducting qubit}

\begin{abstract}
We demonstrate amplification of a microwave signal by a strongly driven two-level system in a coplanar waveguide resonator. The effect known from optics as dressed-state lasing is observed with a single quantum system formed by a persistent current (flux) qubit. The transmission through the resonator is enhanced when the Rabi frequency of the driven qubit is tuned into resonance with one of the resonator modes. Amplification as well as linewidth narrowing of a weak probe signal has been observed. The laser emission at the resonator's fundamental mode has been studied by measuring the emission spectrum. We analyzed our system and found an excellent agreement between the experimental results and the theoretical predictions obtained in the dressed-state model.

\end{abstract}

\date{\today}
\author{G.~Oelsner}
\affiliation{Institute of Photonic Technology, P.O. Box 100239, D-07702 Jena, Germany}
\author{P.~Macha}
\affiliation{Institute of Photonic Technology, P.O. Box 100239, D-07702 Jena, Germany}
\author{O.V.~Astafiev}
\affiliation{NEC Smart Energy Research Laboratories. Tsukuba, Ibaraki, 305-8501, Japan}
\author{E.~Il'ichev}
\affiliation{Institute of Photonic Technology, P.O. Box 100239, D-07702 Jena, Germany}
\author{M.~Grajcar}
\affiliation{Department of Experimental Physics, Comenius University, SK-84248 Bratislava,
Slovakia}
\author{U.~H\"ubner}
\affiliation{Institute of Photonic Technology, P.O. Box 100239, D-07702 Jena, Germany}
\author{B.I.~Ivanov}
\affiliation{Institute of Photonic Technology, P.O. Box 100239, D-07702 Jena, Germany}
\author{P.~Neilinger}
\affiliation{Department of Experimental Physics, Comenius University, SK-84248 Bratislava,
Slovakia}
\author{H.-G.~Meyer}
\affiliation{Institute of Photonic Technology, P.O. Box 100239, D-07702 Jena, Germany}

\maketitle

The concept of inversionless amplification of a probe field by
coherently strongly driven two-level systems has
been pioneered several decades ago \cite{Rautian, Mollow, Haroche72}. The coupling between the
driving field and the two-level quantum system is characterized by
the so-called Rabi frequency. Two resonances occur at the Rabi sidebands
with the frequencies \cite{Mompart2000}
\begin{equation}
\omega_s=\omega_d\pm\Omega_R,	
	\label{Eq:probe}
\end{equation}
where $\Omega_R=\sqrt{\Omega_{R0}^2+\delta^2}$ is the
generalized Rabi frequency at the detuning $\delta=\omega_d-\omega_q$ between driving
field frequency $\omega_d$ and qubit transition frequency $\omega_q$, and $\Omega_{R0}$ is the
on-resonance Rabi frequency. The analysis of such systems shows that there is an absorption of the probe
field at one of the Rabi sidebands ($\omega_p=\omega_d + \Omega_R$) and a gain at the other one
($\omega_p=\omega_d - \Omega_R$). Indeed, the amplification (damping) was realized experimentally by Wu
\emph{et al.} \cite{Wu}, and lasing was demonstrated by Khitrova \emph{et al.} \cite{Khitrova}.

The active medium of conventional lasers and amplifiers consists of many quantum systems,
which are usually natural molecules and atoms. Due to the tiny size of these objects, they are
weakly coupled to the cavity.
Nevertheless, the strong coupling regime has been achieved and a single atom laser with vanishing pumping threshold has been convincingly demonstrated \cite{McKeever}.
By using superconducting qubits as artificial atoms this regime can be achieved easily.
As a consequence, lasing action with a single Josephson-junction charge qubit
has been realized \cite{Astafiev2007}.

Characteristic frequencies of superconducting qubits belong to the microwave
frequency range. Recent activities, motivated by quantum limited measurements at
this range and the implementation of quantum information processing
devices, require extremely low-noise microwave amplifiers. Such amplifiers are
usually based on the nonlinearity of superconducting weak links
\cite{Beltran2008, Yamamoto2008}. Recently, the squeezing of quantum noise
and its measurement below the standard quantum limit was demonstrated
\cite{Mallet2011}.  These successful experiments motivated the development of
a new generation of parametric amplifiers based on 'classical' Josephson junction
structures \cite{Bergeal2010, Hatrige2011}.  However, quantized energy levels
of superconducting qubits can also be used for these purposes.
For instance, the stimulated emission by a superconducting qubit was successfully employed for the amplification of a microwave signal passing through a transmission line
\cite{Astafiev2010}.

In this Letter, we show that the concept of dressed-state lasers \cite{Agarwal90,
Mompart2000} can be used for the amplification of a microwave signal. We
demonstrate experimentally a proof of principle of a
dressed-state amplifier with a single two-level system and provide a full theoretical analysis.  The present
realization exploits the direct transition at the Rabi frequency \cite{Hauss2008} $\Omega_R$ rather
than the sideband transitions $\omega_s$, usually used in quantum optics. Here the Rabi
frequency is tuned into resonance with the oscillator, providing a qubit-resonator
energy exchange at the Rabi frequency.

\begin{figure}[th]
\begin{minipage}{4 cm}
\includegraphics[width=4 cm,angle=0]{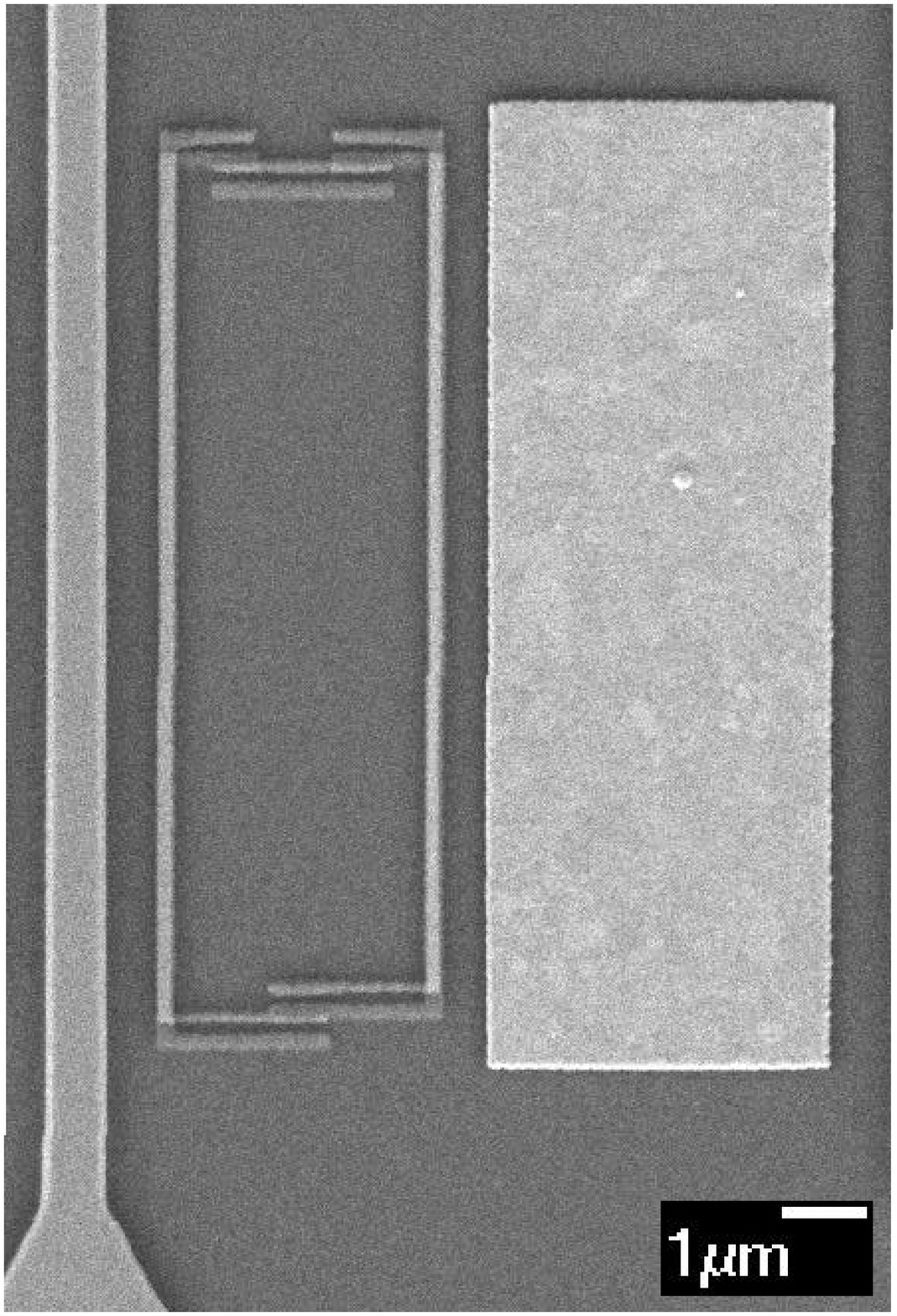}
\end{minipage}
\begin{minipage}{4 cm}
\includegraphics[width=4 cm,angle=0]{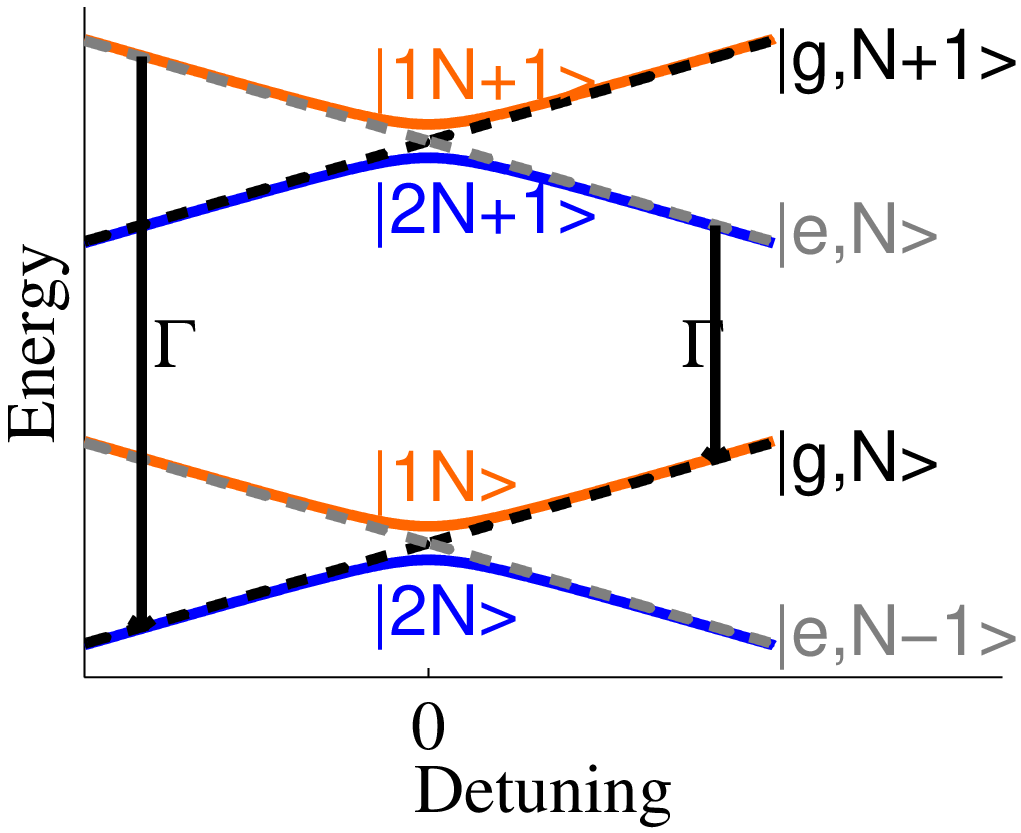}
\includegraphics[width=4 cm,angle=0]{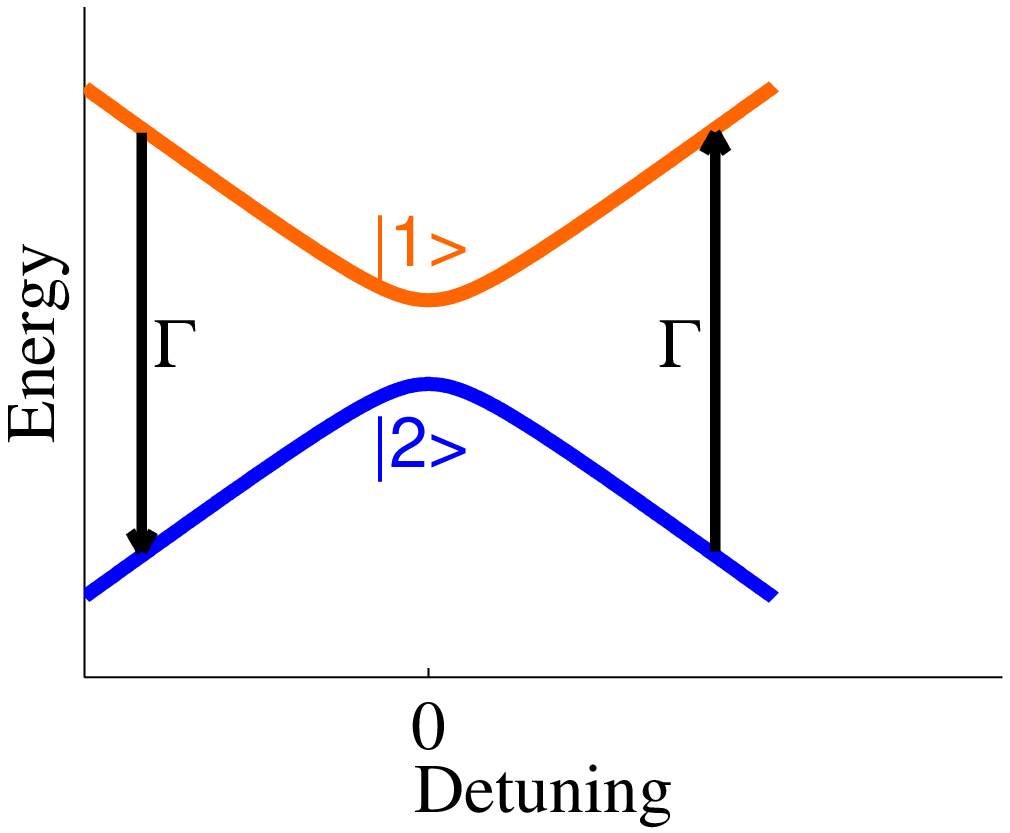}
\end{minipage}
\begin{picture}(200,0)
\put(-15,7){\large (a)}
\put(110,97){\large (b)}
\put(110,7){\large (c)}
\end{picture}
\caption{(color online) (a) Electron micrograph of the central part of the resonator:
the central line of the resonator (left), qubit (middle), and the gold film resistor (right). (b) Sketch of the  dressed system manifolds N and N+1. The dashed lines correspond to the uncoupled states. The qubit relaxation with rate $\Gamma$ is sketched for the uncoupled states. When coupled the levels will split depending on the photon state $N$. (c) After tracing over the photon number N an effective two-level system, denoted with states $\ket{1}$ and $\ket{2}$ is obtained. Both the sign and strength of the relaxation (excitation) in this system depend on the detuning. For large detunings $\left| \delta\right|$ the excitation (relaxation) rate can be identified with the qubit relaxation rate (see Eq.~\ref{Eq:dressed_relax}). This feature follows directly from the property of the superposition between qubit ground and excited state. }
\label{Fig:qubit}
\end{figure}

In order to show dressed-state amplification we have chosen a flux (or
persistent-current qubit) as an artificial atom coupled to a superconducting
coplanar waveguide resonator. The  niobium (Nb) resonator was fabricated by e-beam
lithography and dry etching of a 200-nm-thick Nb film, deposited on a silicon
substrate.  The length of the resonator's central conductor is $L\simeq$ 23
mm which results in a resonance frequency of $\omega_r/2\pi \approx 2.5$ GHz for the
fundamental half-wavelength mode.  The width of the central conductor is 50 $\mu$m, and the gap between the central conductor and the ground plane is
30 $\mu$m resulting in a wave impedance of about 50 $\Omega$. In the middle of the
resonator the central conductor is tapered to a width of 1 $\mu$m for a length of
30 $\mu$m with 9 $\mu$m gap (see Fig. \ref{Fig:qubit}), which provides better
qubit-resonator coupling and a small impedance mismatch to detune the harmonics
of the resonator. We achieved a detuning of about 25 linewidths between fundamental mode and third harmonic. At millikelvin temperatures the resonator quality factor is
measured to be $Q\simeq$ 4 $\times$ 10$^4$.

The aluminum flux qubit was fabricated in the central part of the resonator by
making use of conventional shadow evaporation technique. The qubit loop size ($12 \times 5$ $\mu \text{m}^2$) is interrupted by three Josephson junctions. Two of them have a nominal
size of ($550 \times 120$ nm$^2$), while the third is about 25 percent smaller.  Two
cryoperm shields and one superconducting shield are used to minimize the influence of
the external magnetic fields. The sample was thermally anchored to the mixing
chamber of a dilution refrigerator, providing a base temperature of about 10 mK.

The parameters of the qubit, the persistent current $I_p$ = 12 nA and the minimum transition frequency (gap) $\Delta/2\pi$ = 3.7 GHz,
were determined from the transmission of the resonator
measured as a function of the qubit transition frequency.
As the gap of the qubit is higher
than the frequency of the fundamental mode, we observe a dispersive shift of the resonator's fundamental frequency.
The energy gap and persistent current of the qubit were separately determined from this dispersive shift \cite{Ili2004}
and from resonant interaction \cite{Born2004, Wallraff2004} combined with the positions of the
anticrossings \cite{Oelsner2010, Omelyanchouk2010}. Similar to Ref.~\cite{Izmalkov2008} both sets of data are completely consistent.

The for a flux qubit unusual low persistent current was chosen for emission stability and efficiency. We have decreased the critical current of the Josephson junctions
and increased the energy difference of the two-level system to make it more immune
against flux fluctuation \cite{You2007}.
Although the qubit became more sensitive to charge fluctuation, we did not observe any significant charge noise effects.
In addition, the relaxation time was intentionally decreased by a
gold resistor placed in parallel to one arm of the qubit loop
(see Fig. \ref{Fig:qubit}).

\begin{figure}[th]
\includegraphics[width=8cm,angle=0]{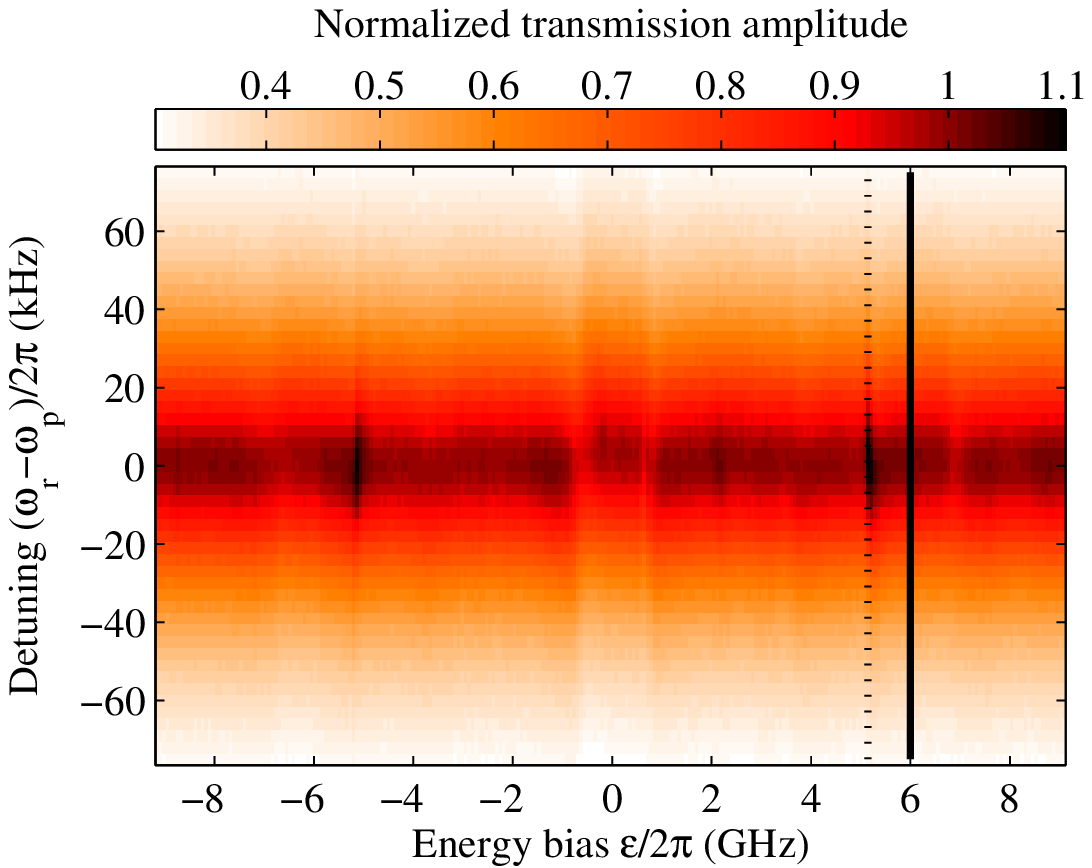}
    \begin{picture}(100,0)    
        \put(30,35){\includegraphics[width=80\unitlength]{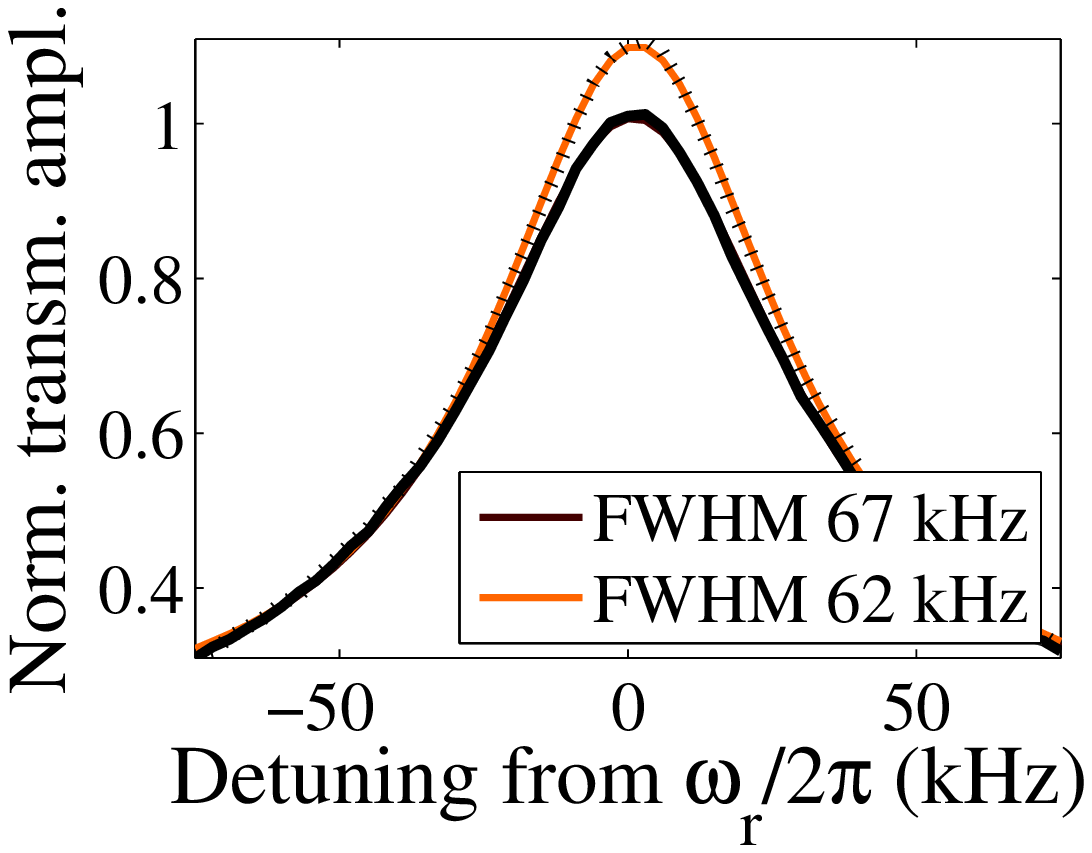}}
    \end{picture}
\includegraphics[width=8cm,angle=0]{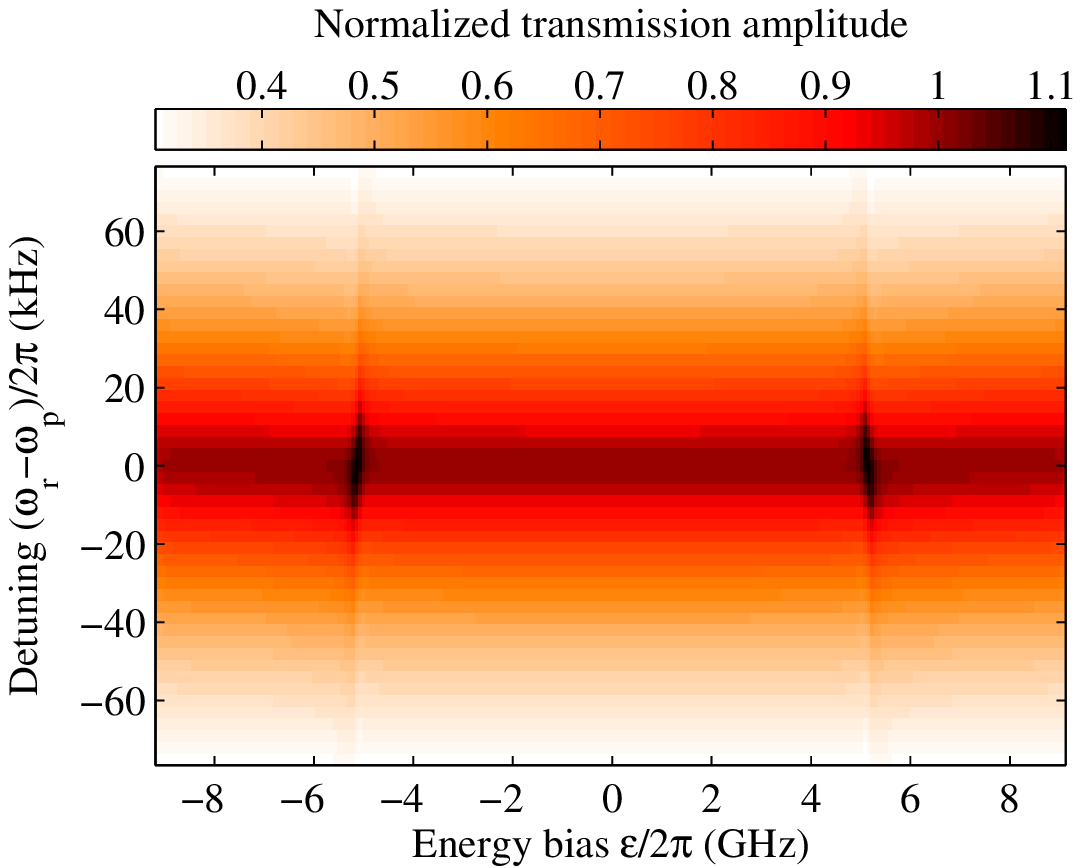}
    \begin{picture}(100,0)    
        \put(30,35){\includegraphics[width=80\unitlength]{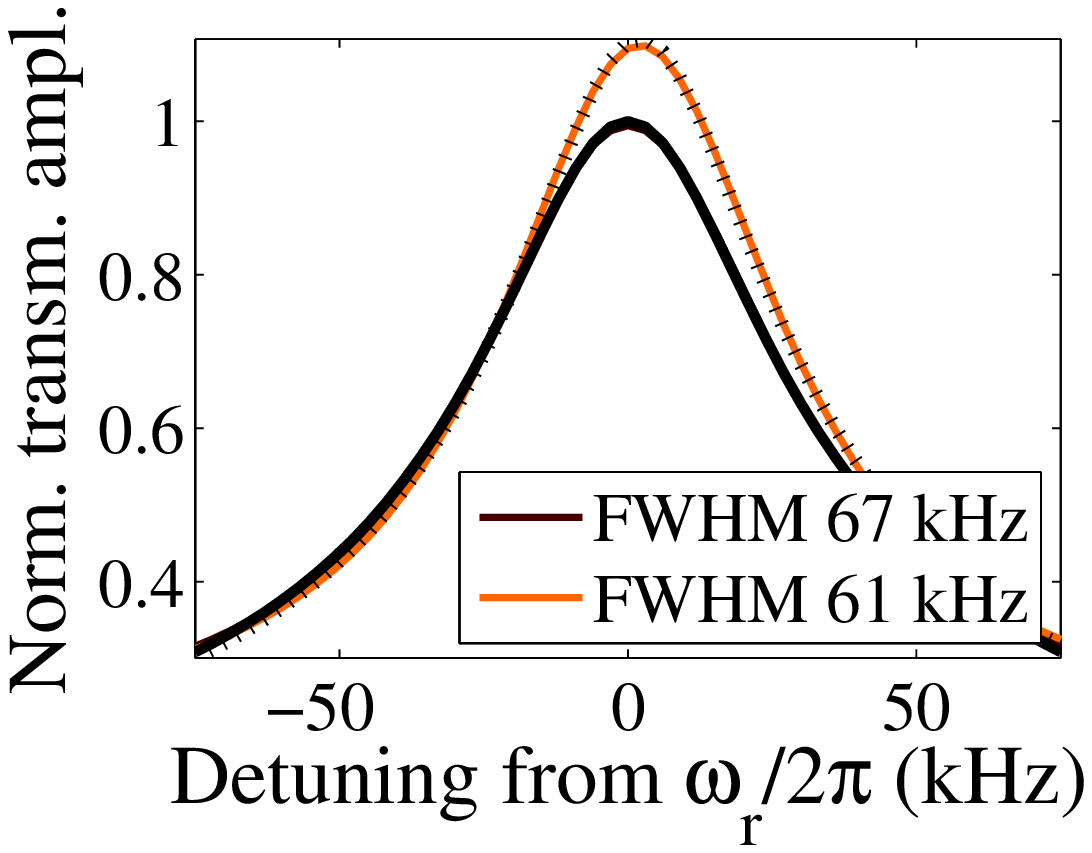}}
    \end{picture}
\begin{picture}(200,0)
\put(20,240){\large (a)}
\put(20,45){\large (b)}
\end{picture}
\caption{(color online) (a)
Normalized transmission amplitude of the probe signal as a function of the
detuning from the resonator's
fundamental mode frequency $(\omega_p - \omega_r)/2\pi$ and qubit energy bias $\epsilon/2\pi$. The inset shows the Lorentzian shaped transmission vs. the detuning in the amplification point (dashed line) and out of the Rabi resonance (solid line). The experiments were done
at the presence of a strong driving field applied at the third harmonic
of the resonator. (b) Numerical simulations by making use of
Eq.~\ref{Eq:Master} in a four photon space. The probing power is assumed to be
weak creating a coherent state with a mean photon number less than one in the fundamental mode.}
\label{Fig:transmission}
\end{figure}

In the first set of experiments, the frequency of the driving signal $\omega_d$
was set to the frequency of the third harmonic of the resonator $\omega_d
\approx 3\omega_r$. The odd harmonics of the resonator provide optimal coupling of
the driving signal to the qubit. The transmission of a weak coherent probe
signal, populating the resonator with an average photon number less than one, was measured at frequencies $\omega_p$ close to
the frequency of the resonator's fundamental mode $\omega_r$.  The dependence of
the transmission on the qubit transition frequency is shown in Fig.~\ref{Fig:transmission}. The
two dark areas correspond to the amplification. As expected this amplification is observed
when the transition frequency of the qubit $\omega_q=\sqrt{\Delta^2+\epsilon^2}$ is tuned to the interaction point, which depends on the driving strength at $\omega_d$ (see below). Here $\hbar \epsilon = \Phi_0 I_p \left( 2\Phi_x / \Phi_0-1 \right)$ is the energy bias of the qubit and $\Phi_0$ is the magnetic flux quantum. $\Phi_x$ is the external flux in the qubit loop.
\begin{figure}[th]
\includegraphics[width=8cm,angle=0]{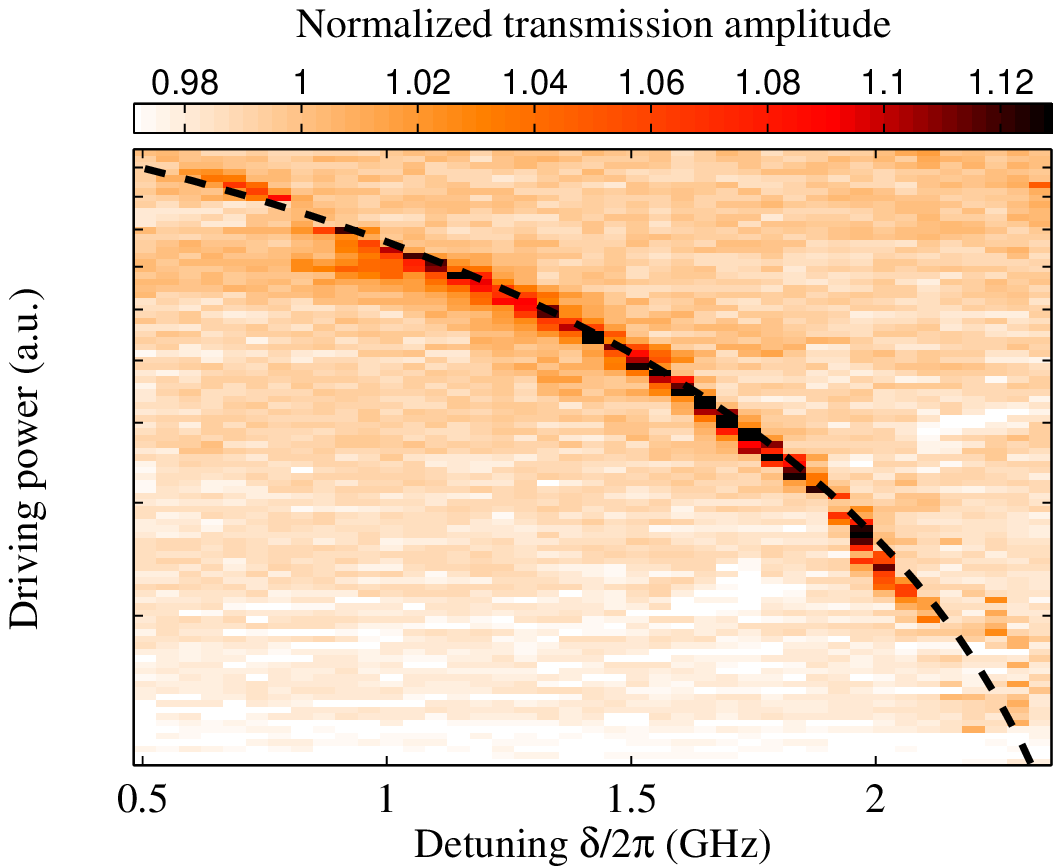}
\includegraphics[width=8cm,angle=0]{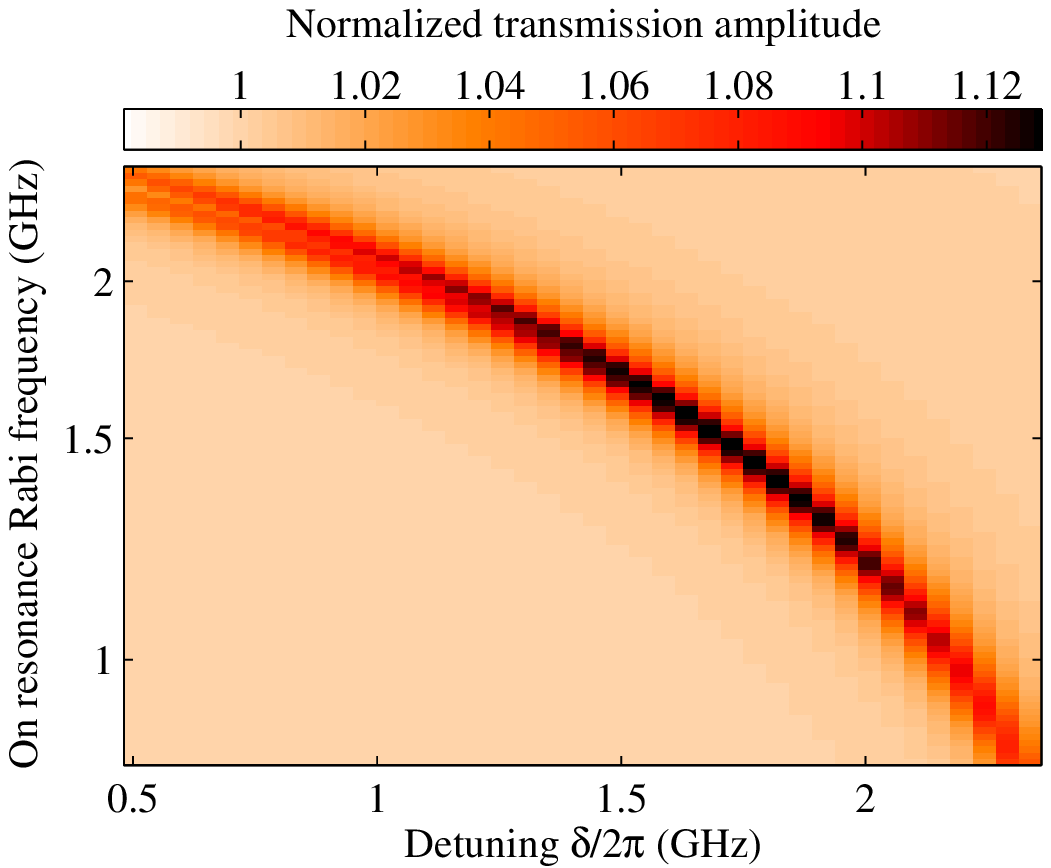}
\begin{picture}(200,0)
\put(20,225){\large (a)}
\put(20,35){\large (b)}
\end{picture}
\caption{(color online) (a) Normalized transmission amplitude as a function of the detuning $\delta$ (between the strong driving $\omega_d$ and the qubit transition frequency
$\omega_q$) and the driving power.
The dashed line is calculated according to Eq. \ref{Eq:Rabifreq} and used to map the on resonance Rabi frequency $\Omega_R^{\langle N \rangle}(\delta=0)$ to the driving power. (b) Numerical simulations of the
experimental results shown in (a) by use of Eq.~\ref{Eq:Master}.}
\label{Fig:powerdep}
\end{figure}
Quantitatively, the amplitudes of the resonance curves at these points are increased by $10$~\% and
show linewidth narrowing; the resonator loss rate is reduced as well by about $10$~\%, see inset in Fig.~\ref{Fig:transmission}.
For quantitative analysis, we consider the Hamiltonian of a driven qubit, using an approach similar to Ref. \cite{Hauss2008}. However, we use a quantized driving field.
The photon field is described by the creation and annihilation operators $b^\dagger$ and $b$; the coupling between the field and the qubit is characterized by the energy $\hbar g_d$. The qubit is represented by the Hamiltonian $H_q = \frac{\hbar \omega_q}{2}\sigma_z$ in the basis of the ground and excited states $|g\rangle$ and $|e\rangle$. $\sigma_{x,y,z}$ are the Pauli matrices.
By neglecting the diagonal coupling term, proportional to $\sigma_z$, as well as the non-resonant terms ($b^\dagger \sigma_+$ and $b \sigma_-$, where $\sigma_\pm = (\sigma_x \pm i\sigma_y$)/2), we arrive at the Jaynes-Cummings Hamiltonian
\begin{align}
H_{d}\!=\!\hbar\omega_d \left( b^\dagger \! b \! + \frac{1}{2} \! \right) \! + \!\hbar\frac{\omega_q}{2} \sigma_z \!+ \! \hbar g_d \! \frac{\Delta}{\omega_q} \left(\!b^\dagger \!\sigma_-\!+ \! b\sigma_+\!\right). \label{Eq:Ham_qubit1_2}
\end{align}
By diagonalization the Hamiltonian is transformed to the dressed-state basis. This yields a ladder
\begin{equation}
H_d= \hbar\omega_d \hat{n} + \frac{\hbar}{2}\hat{\Omega}_R,
\end{equation}
where we introduced the number operator $\hat{n} = \sum_N N(|1 N\rangle\langle 1 N| + |2 N\rangle\langle 2 N|)$ and the Rabi operator $\hat{\Omega}_R=\sum_N \Omega_R^{(N)}(|1 N\rangle\langle 1 N| - |2 N\rangle\langle 2 N|)$. The dressed states $|1 N\rangle $ and $|2 N\rangle$ are found by a standard transformation \cite{CohenTannoudji}.
We use the modified generalized Rabi frequency as:
\begin{align}
\Omega_R^{(N)} =\sqrt{\delta^2 + 4g_d^2 N \Delta^2/\omega_q^2} \label{Eq:Rabifreq},
\end{align}
where $\delta=\omega_d-\omega_q$ is the detuning.
For large $N$ and small deviations from the average photon number $\langle N \rangle$ of the driving cavity field, $\Omega_R^{(N)}$ can be substituted by the constant value $\Omega_R=\Omega_R^{(\langle N \rangle)}$.
In the dressed-state basis the interaction between the dressed-state system and the resonator's fundamental
mode is described by the Hamiltonian:
\begin{align}
H\!=\!H_d \! + \! \hbar\omega_r a^\dagger a \! + \! \hbar g_e \left( \! \frac{\Omega_p}{\Omega_R}\! \sigma_x \! + \! \frac{\delta}{\Omega_R}\! \sigma_z \! \right)\!\left(\! a^\dagger \! + \! a \! \right)\! +\! H_p . \label{Eq:Ham_final}
\end{align}
Here, the fundamental mode of the resonator is expressed by the product of the creation and annihilation operators $a^\dagger$ and $a$; $g_e$ is the effective coupling constant of that mode to the qubit, and the probing field is described by
$H_p = \hbar \Omega_p \cos \omega_p t \left( a^\dagger + a \right)$. Note that the Hamiltonian in Eq.~(\ref{Eq:Ham_final}) reflects the dynamics at the Rabi frequency: the transitions between the dressed states from different manifold $N$ are neglected.

The dynamics of this Rabi lasing system is essentially dissipative. The effects of dissipation can be accounted for
by the Liouville equation for the density matrix of the system, including the relevant damping terms, which are assumed to be of Markovian form. The corresponding damping (so-called Lindblad) terms are found for relaxation with rate $\Gamma$ and dephasing with rate $\Gamma_\varphi$ of the qubit as well as the photon decay of the resonator with rate $\kappa$.

The Liouville equation for the density matrix of the system can be written in the usual form
\begin{align}
-\frac{i}{\hbar} \left[ H,\rho \right] + L^R (\rho) + L_r (\rho)= 0 \label{Eq:Master}
\end{align}
for the reduced elements of the density matrix $\rho$ in the rotating frame.
The damping terms, $L^R$ from the qubit and $L_r$ for the resonator, are written in the dressed-state basis. Since the operators $a^\dagger$ and $a$ are the same in both basis, the photon decay of the resonator's fundamental mode is not changed after this transformation.
Nevertheless, the population of the dressed states itself is modified by the relaxation rates of the qubit. By omitting the qubits decoherence rate and the oscillating terms in the rotating frame the relaxation of the qubit in the dressed basis reads
\begin{align}
L^R_{11}&= \frac{\Gamma}{2} \frac{\delta}{\Omega_R} \left[ \left( 1-\frac{\delta}{\Omega_R} \right) \rho_{11} + \left( 1+\frac{\delta}{\Omega_R} \right) \rho_{22} \right] \nonumber \\
L^R_{22}&= -L^R_{11}& \label{Eq:dressed_relax}\;.
\end{align}
Here, $L^R_{11}=tr^{(N)} \langle 1 N|L_q|1 N\rangle$ and $L^R_{22}=tr^{(N)} \langle 2 N|L_q|2 N\rangle$ are two reduced elements of the Lindblad operator for the dressed-state system, which yield from the standard Markovian Lindblad operator $L_q$ in the qubit's eigenbasis. We similarly introduced the reduced elements of the density matrix $\rho_{11}=tr^{(N)} \langle 1 N|\rho|1 N\rangle$ and $\rho_{22}=tr^{(N)} \langle 2 N|\rho|2 N\rangle$.
Since the states of this effective two-level system are formed by a superposition of qubit ground and excited state, the rate of relaxation depends on the weight of both states in this superposition. This is illustrated in Fig.\ref{Fig:qubit}~(b) and (c). The relaxation is mainly defined by the detuning. Moreover, for positive detuning $\delta$ the relaxation of the qubit puts the effective two-level system to the higher energetic state. Therefore, in the frame of the dressed-state model, a population inversion of the energy levels is created. For small detunings the dephasing rate equalizes the population of the two reduced dressed levels.

In order to find the quasi steady state density matrix elements we solved Eq.~\ref{Eq:Master} numerically. We limited the photon space of the resonators fundamental mode to a size of four, since we considered weak probing but expect an increasing photon number due to the amplification effect. The transmission of the resonator is then given by the expectation value of the annihilation operator $ \langle a\rangle$ \cite{Omelyanchouk2010}.
Theory and experimentally obtained transmission data are in good agreement (Fig.\ref{Fig:transmission}). We observe the amplification of the weak probe signal as well as the linewidth narrowing, see Fig.~\ref{Fig:transmission}.  From the best fitting the following parameters of the system were obtained $g/2\pi \;\approx$ 0.8~MHz, $\kappa/2\pi$ = 65~kHz, $\Gamma/2\pi \approx 80$~MHz, and $\Gamma_\varphi \ll \Gamma$. The pure dephasing is negligible relative to the relaxation rate which was intensionally enhanced by the gold film resistor.

The power dependence of the amplification point (Fig.~\ref{Fig:powerdep}) has been described as well. The dotted line in Fig.~\ref{Fig:powerdep}(a) keeps the resonance condition $\Omega_R = \omega_p$ as a function of the detuning $\delta$ and driving power. By making use of this condition the photon number was calculated according to Eq.~\ref{Eq:Rabifreq}. Consistence between theory and experiment was also found in this case. The optimum of the amplification process results from the dependence of the effective excitation rates and the coupling constant on the detuning $\delta$, since a monotonic dependence of the qubit excitation rate ($\Gamma\propto \delta / \Omega_R$) is compensated by the decreasing of the effective coupling  ($g_{e}\propto \Omega_d / \Omega_R$).
\begin{figure}[th]
\includegraphics[width=8cm,angle=0]{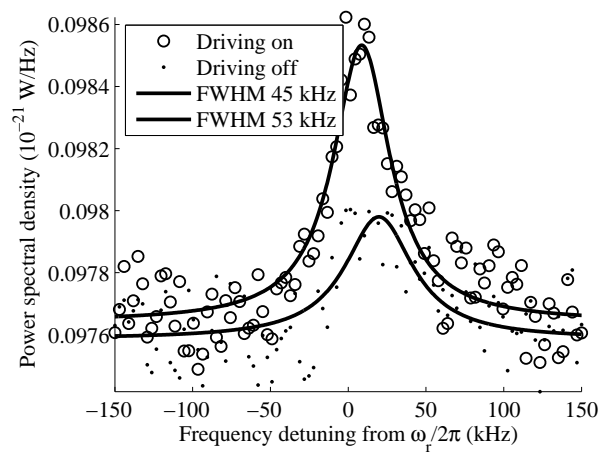}
\caption{(color online) Spectral emission from the resonator. The solid lines correspond to best Lorentzian fits of the measured data and are used to reconstruct the linewidths. The power spectral density measured without driving field (dots) corresponds to the thermal occupation of the resonator at an effective temperature of 30 mK. The background yields from the noise of the cold amplifier, with an value of 7 K. If the lasing process is turned on by driving the resonators 3rd. harmonic we observe a clear increase of the emission (open circles). This strong increase in emission and the lower linewidth are signs of a self oscillation of the resonator.}
\label{Fig:emmission}
\end{figure}

In a second set of experiments we removed the probing field and measured the emission of the resonator at its fundamental mode with optimum amplification parameters. The results are shown in Fig~4. An increasing of the emission in the presence of strong driving (open circles) was clearly observed, compared to the thermal response of the resonator (dots). Standard fitting of the data by Lorentzian curves demonstrates a linewidth narrowing. In general, this could indicate laser action in the system, but this issue requires further study

In conclusion, we designed a superconducting resonator-qubit system and demonstrated experimentally the main features of dressed-state amplification of a probe field by a strongly driven qubit directly at the Rabi-frequency. Linewidth narrowing for the transmission of the weak probe field was observed. Furthermore, free emission provided by the strong driving was detected. Numerical simulations done in the frame of the dressed-state model and obtained experimental results are completely consistent. We believe that optimizations in the design could lead to the realization of new types of quantum microwave amplifiers and sources.

The authors gratefully acknowledge the financial support of the EU through the project SOLID,  IQIT and  ERDF OP R\&D,
Project CE QUTE \& metaQUTE. M.G. and P.N. were supported  by the
Slovak Research and Development Agency under the contract
No. APVV-0432-07, APVV-0515-10 and LPP-0158-09. G.O. thanks Ya. S. Greenberg and S. N. Shevchenko for helpful discussions.

\end{document}